\documentclass[pdflatex,sn-mathphys-num]{sn-jnl}

\usepackage{graphicx}%
\usepackage{multirow}%
\usepackage{amsmath,amssymb,amsfonts}%
\usepackage{amsthm}%
\usepackage{mathrsfs}%
\usepackage[title]{appendix}%
\usepackage{xcolor}%
\usepackage{textcomp}%
\usepackage{manyfoot}%
\usepackage{booktabs}%
\usepackage{algorithm}%
\usepackage{algorithmicx}%
\usepackage{algpseudocode}%
\usepackage{listings}%
\usepackage{siunitx}%

\theoremstyle{thmstyleone}%
\theoremstyle{thmstyletwo}%

\theoremstyle{thmstylethree}%

\begin{document}

\title[Generation and detection of squeezed states via a synchronously pumped optical parametric oscillator]{Generation and detection of squeezed states via a synchronously pumped optical parametric oscillator}

\author*[1,2]{\fnm{Edoardo} \sur{Suerra}}\email{edoardo.suerra@unimi.it}
\author[1,2]{\fnm{Samuele} \sur{Altilia}}
\author[1,2]{\fnm{Stefano} \sur{Olivares}}
\author[1,2]{\fnm{Alessandro} \sur{Ferraro}}
\author[3]{\fnm{Francesco} \sur{Canella}}
\author[4,2]{\fnm{Dario} \sur{Giannotti}}
\author[3]{\fnm{Gianluca} \sur{Galzerano}}
\author[1]{\fnm{Sebastiano} \sur{Corli}}
\author[1,2,3]{\fnm{Enrico} \sur{Prati}}
\author[1,2]{\fnm{Simone} \sur{Cialdi}}

\affil[1]{\orgdiv{Dipartimento di Fisica}, \orgname{Università degli Studi di Milano}, 
\orgaddress{\street{Via Celoria 16}, \city{Milan}, \postcode{I-20133}, \country{Italy}}}
\affil[2]{\orgname{Istituto Nazionale di Fisica Nucleare (INFN), Sezione di Milano}, 
\orgaddress{\street{Via Celoria 16}, \city{Milan}, \postcode{I-20133}, \country{Italy}}}
\affil[3]{\orgname{Istituto di Fotonica e Nanotecnologie – CNR}, 
\orgaddress{\street{Piazza Leonardo da Vinci 32}, \city{Milan}, \postcode{20133}, \country{Italy}}}
\affil[4]{\orgdiv{Dipartimento di Fisica}, \orgname{Politecnico di Milano}, 
\orgaddress{\street{Piazza Leonardo da Vinci 32}, \city{Milan}, \postcode{20133}, \country{Italy}}}

\abstract{
A synchronously pumped optical parametric oscillator (SPOPO) operating at \SI{93}{\mega\hertz} is used to generate squeezed states at \SI{1035}{\nano\meter}.
The system features a counter-propagating beam at the same wavelength as the quantum state, which simultaneously actively stabilizes the cavity and, after transmission, acts as the local oscillator for homodyne detection.
By deriving the local oscillator directly from the SPOPO cavity, the setup establishes an intrinsically excellent spatial mode overlap and high interference visibility, forming a distinctive self-referenced architecture.
Two spatial light modulators enable precise spectral shaping of both the pump and the local oscillator in amplitude and phase, allowing investigation of the spectral properties of the generated states.
The versatility of the setup further allows exploration of different SPOPO configurations, including regimes with varied finesse and escape efficiency.
Representative measurements, including homodyne traces and squeezing levels as functions of pump power and local oscillator bandwidth, demonstrate the performance of the system.
Theoretical simulations based on a multimode singular-value-decomposition model reproduce well the measured dependence of squeezing on pump power and LO bandwidth, confirming the accuracy of the description and the robustness of the setup.
Measured squeezing levels up to \SI{-3.3}{\decibel} are achieved, corresponding to \SI{-5.7}{\decibel} at SPOPO output, evidencing the robustness and versatility of this platform for stable pulsed squeezed-light generation and advanced quantum optical applications.
}




\maketitle

\section{Introduction}\label{sec:intro}
Squeezed states of light are a fundamental resource for a broad range of quantum technologies, including quantum communication, quantum computation, quantum metrology, and quantum networks \cite{Braunstein2005,Weedbrook2012}. 
By reducing quantum noise below the shot-noise limit in a given quadrature, they enable enhanced sensitivity and constitute a versatile platform for encoding and processing quantum information. 
While continuous-wave (CW) squeezed light has been extensively studied and successfully applied in quantum optics experiments \cite{Lvovsky2015,Vahlbruch2016}, the generation of pulsed squeezed states has attracted growing attention in recent years. 
Pulsed systems naturally provide access to a rich multimode temporal and spectral structure, which is well suited for high-dimensional quantum information protocols \cite{Fabre2020,Roslund2014}. 
Moreover, pulsed squeezed states are compatible with time and frequency multiplexing schemes, making them highly promising for scalable quantum networks and quantum memories \cite{Reid2009,MedeirosdeAraujo2014}. 
Also, temporal single-mode operation is equally important and desirable, as single-mode sources are a key ingredient for protocols that rely on well-defined Gaussian states, such as Gaussian boson sampling, and represent an essential resource for photonic quantum computing and simulation \cite{Hamilton2017}. 
High-performance generation of squeezed light in pulsed regimes typically hinges on precise control of cavity parameters and adequate spectral filtering of the output field; mastering these elements is a milestone for pulsed squeezing platforms.
Several platforms have been developed to generate pulsed squeezed states, including nonlinear waveguides, optical fibers, and optical parametric oscillators (OPOs) \cite{Wasilewski2006}. 
Among these, synchronously pumped optical parametric oscillators (SPOPOs) stand out for their ability to combine efficient nonlinear interaction with cavity enhancement, allowing strong squeezing over broad spectral bandwidths \cite{Eckstein2011,Patera2010}. 
However, the implementation of both OPOs and SPOPOs presents important experimental challenges. 
In particular, robust cavity stabilization and efficient mode matching between the generated squeezed field and the local oscillator for homodyne detection are technically demanding, a critical aspect also in applications where OPOs are employed as versatile tools for the mitigation of phase noise and diffusion in optical fields \cite{Cialdi2020,Notarnicola2022}.
Furthermore, dispersion management and spectral control are often critical for maintaining high levels of squeezing and for tailoring the multimode structure of the output states \cite{Averchenko2024,Kouadou2023}.
Different stabilization strategies and detection schemes have been proposed to address these requirements \cite{Slusher1985,Ou1992,Roslund2014,Vahlbruch2016,Cialdi2021}, but they often involve trade-offs in terms of robustness, flexibility, or achievable visibility in homodyne detection. 
Ensuring excellent spatial overlap between the local oscillator and the squeezed field is especially challenging, yet crucial for accessing the full quantum properties of the generated state \cite{Breitenbach1997,Lvovsky2015}.

Here, we present an experimental setup that addresses these limitations by introducing a novel approach to local oscillator generation. 
Our scheme is based on a SPOPO operating at \SI{93}{\mega\hertz} and \SI{1035}{\nano\meter}, which is actively stabilized using a beam counter-propagating with respect to the quantum state at the same wavelength. 
Crucially, such beam, after transmission through the cavity, serves as the local oscillator (LO) for homodyne detection. 
By deriving the LO directly from the cavity, the system naturally provides an excellent spatial mode overlap and high visibility, a key advantage over conventional external LO schemes.
In addition, we integrate two spatial light modulators (SLMs) into the setup, enabling precise amplitude and phase shaping of both the pump and the local oscillator. 
This versatility allows us to explore different SPOPO configurations by varying cavity finesse, escape efficiency, and dispersion. 
Here, we provide a general overview of the setup and present representative measurements, i.e., homodyne traces and squeezing levels as a function of pump power and local oscillator spectral bandwidth. 
In addition, we compare the experimental results with theoretical simulations based on a multimode singular-value-decomposition model, which accurately reproduce the observed trends of squeezing level, confirming the validity of the theoretical framework.
We report levels of measured squeezing up to \SI{-3.3}{\decibel}, demonstrating both the robustness of the stabilization scheme and the flexibility of the platform. 
This work thus establishes a solid experimental basis for future developments in multimode quantum optics and quantum information processing.

\section{Experimental setup}\label{sec:setup}
\begin{figure}
    \centering
    \includegraphics[width=10cm]{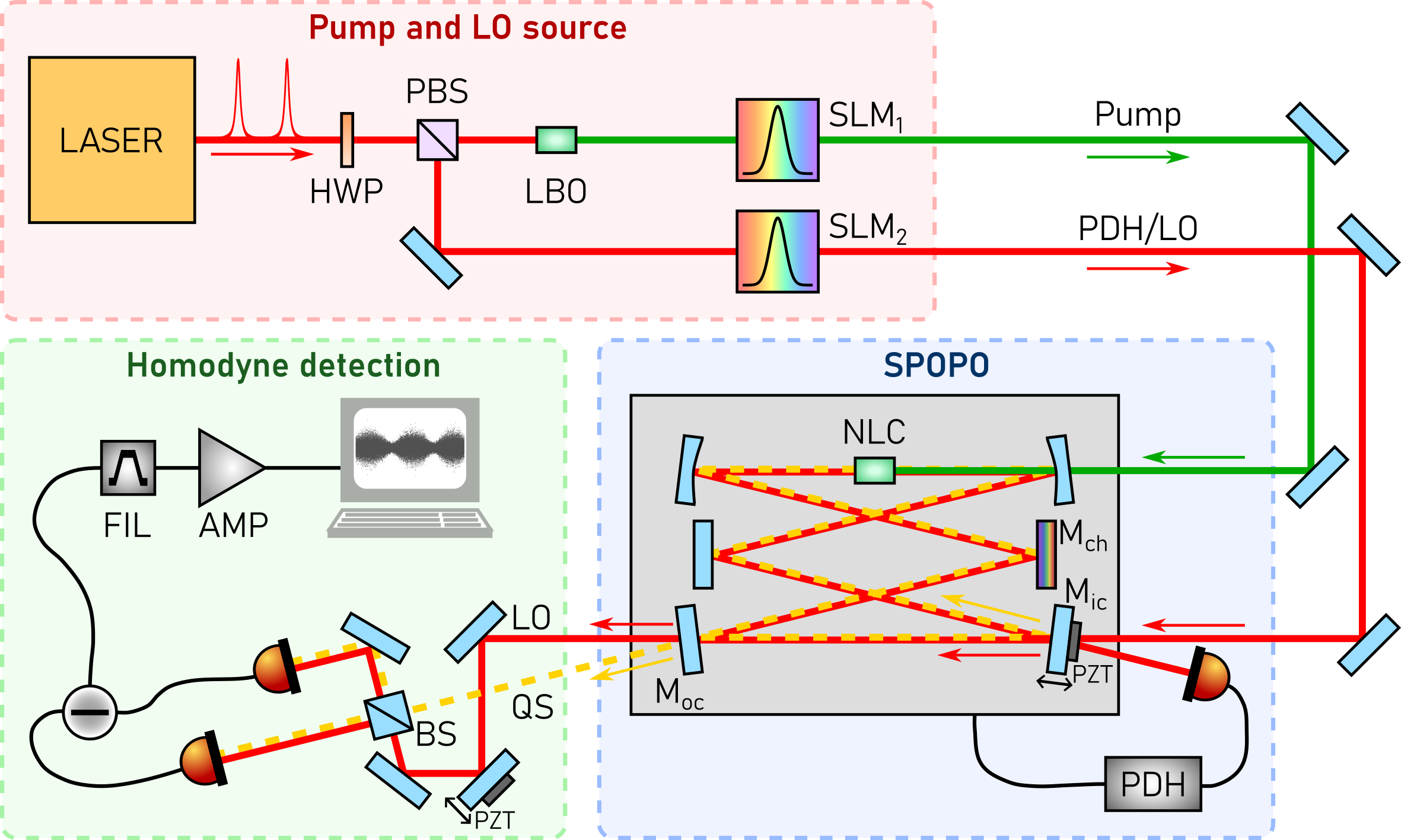}
    \caption{
    Scheme of the experimental setup.
    Pump and LO are generated from a \SI{93}{\mega\hertz} single mode-locked laser, and independently spectrally shaped using two SLMs.
    The SPOPO is frequency-stabilized to the laser repetition rate, and both the quantum state and the LO originate from the same cavity mode.
    A balanced homodyne detection stage measures the frequency-resolved squeezing level.
    HWP half-waveplate, PBS polarizing beam splitter, LBO lithium-borate crystal, SLM$_\mathrm{i}$ spatial light modulators, PDH Pound-Drever-Hall electronics, M$_\mathrm{ic}$ and M$_\mathrm{oc}$ input and output couplers, M$_\mathrm{ch}$ chirped mirror, NLC nonlinear crystal, LO local oscillator, QS quantum state, PZT piezoelectric actuator, BS 50:50 beam splitter, FIL bandpass filter, AMP differential amplifier.
    }
    \label{fig:setup}
\end{figure}

The setup used for the generation and detection of pulsed squeezed quantum states is illustrated in Figure \ref{fig:setup}, and consists of three main parts: a mode-locked laser source used for both pump and local oscillator radiation, a SPOPO, and a balanced homodyne detection system.
In the said setup, the local oscillator for homodyne detection is directly derived from the SPOPO cavity. 
Using the cavity-transmitted beam as the LO provides a near-perfect spatial mode overlap with the squeezed quantum state. 
When combined with independent spectral shaping of both pump and LO beams and a versatile cavity design, this approach offers a robust and flexible platform for the generation and characterization of pulsed squeezed states.

The source consists of a commercial mode-locked Ytterbium-doped fiber laser with a repetition rate of \SI{93}{\mega\hertz} and a central wavelength of \SI{1035}{\nano\meter}.
Its output is split into two arms: one is frequency-doubled using a \SI{5.0}{\milli\meter}-thick LBO crystal in near non-critical configuration to generate \SI{517.5}{\nano\meter} green pump pulses for the SPOPO; the other arm is used both for frequency locking of the SPOPO and, after transmission through the cavity, as the LO for homodyne detection. 
Both IR and green beams are independently shaped in amplitude and phase using two spatial light modulators (SLM$_1$ and SLM$_2$), placed in the Fourier plane of a folded 4-f system, in a manner similar to that described in \cite{Monmayrant2010}.
This enables shaping the pump spectrum into a Gaussian profile with flat phase and FWHM of \SI{1.0}{\nano\meter}, yielding close-to-transform-limited pulses of \SI{370}{\femto\second} FWHM duration.
The IR arm has a spectral width up to \SI{3}{\nano\meter} FWHM, and a pulse duration of approximately \SI{500}{\femto\second}, also close to transform-limited. 
The selection of this spectral bandwidth is dictated by spatially dependent inhomogeneities in the laser spectrum, as discussed in \cite{Suerra2025}, which require isolating a narrower spectral region of the original beam using the SLMs.

The SPOPO cavity has a total length of \SI{3.23}{\meter}, and consists of six mirrors in a near-confocal geometry, comprising four flat and two curved mirrors with a radius of curvature of \SI{500}{\milli\meter}. 
At the pump wavelength of \SI{517.5}{\nano\meter} all the mirrors have high transmittivity.
At \SI{1035}{\nano\meter} the input coupler has a reflectivity of \SI{99}{\percent}, the output coupler is initially chosen with \SI{81}{\percent} reflectivity, while all other mirrors have high reflectivity (HR).
For this configuration ($R_\mathrm{oc} = \SI{81}{\percent}$) we measured a cavity finesse of $F=24$ with the technique described in \cite{Galzerano2020}.
A \SI{3.0}{\milli\meter}-thick lithium niobate nonlinear crystal is positioned at the cavity waist between the curved mirrors, where the mode radius is approximately \SI{80}{\micro\meter}. 
The pump beam is focused onto the crystal with a waist of \SI{50}{\micro\meter} and adjustable power up to \SI{80}{\milli\watt}.
The SPOPO is frequency-locked to the source laser using the IR beam from SLM$_2$ exploiting the Pound-Drever-Hall technique \cite{Drever1983}, with a modulation frequency of \SI{10}{\mega\hertz}.
The carrier–envelope offset of the laser is adjusted to match the laser comb with the cavity comb \cite{Holzberger2015}.
Our scheme employs a counter-propagating auxiliary beam at the signal wavelength, which naturally separates the beam used for cavity locking from the squeezed output. 
This auxiliary beam also operates as the LO for homodyne detection after transmission through the cavity, providing a near-ideal spatial mode overlap with the quantum state, as they derive from the same spatial mode of the SPOPO. 
The cavity design is highly versatile: mirror reflectivities and intracavity dispersion can be adjusted to achieve different combinations of finesse, escape efficiency, and group-delay dispersion (GDD). 
Dispersion control is implemented by replacing one of the flat high-reflectivity (HR) mirrors with either a standard or a chirped HR mirror. 
In this work, the dispersion introduced by the \SI{3.0}{\milli\meter}-thick LNB crystal ($\sim\SI{850}{\femto\second\squared}$) and the air path ($\sim\SI{50}{\femto\second\squared}$) was compensated with a chirped HR mirror providing \SI{-900}{\femto\second\squared}, yielding a residual GDD close to zero.

Quantum states are detected using a standard frequency-resolved homodyne detection scheme \cite{Bachor2019}: the quantum state and the LO are mixed on a 50:50 beam splitter, and the outputs are detected by a balanced photodiode pair. 
The difference signal is amplified and filtered with a bandpass filter between \SI{400}{\kilo\hertz} and \SI{600}{\kilo\hertz}, which lies well within the SPOPO bandwidth of \SI{3.9}{\mega\hertz} and above mechanical and technical noises. 
As mentioned above, the same IR beam used for SPOPO stabilization serves as the LO after transmission through the cavity, ensuring excellent spatial mode matching with the squeezed state. 
As a result, we routinely achieve high and stable homodyne visibility $\mathrm{vis} \geq \SI{94}{\percent}$. 
A delay line is used to temporally synchronize the LO and signal pulses using a piezoelectric actuator attached on a mirror.

To quantify the actual squeezing level generated by the SPOPO, the measured variance $\sigma^2_\mathrm{hom}$ must be corrected for the total detection efficiency $\eta_\mathrm{hom}$ as (we set to 1 the shot-noise variance) \cite{Leonhardt1995,Olivares21}
\begin{equation}\label{HD:var}
    \sigma^2_\mathrm{hom}(\theta) = 
    1 - \eta_\mathrm{hom}\left[1 - \sigma^2_\mathrm{SPOPO}(\theta)\right] 
    \, ,
\end{equation}
where $\sigma^2_\mathrm{SPOPO}(\theta)$ denotes the actual variance of the SPOPO output state for a given quadrature $X(\theta)$. 
The overall detection efficiency is expressed as the product of several independent contributions \cite{Roman-Rodriguez2024}:
\begin{equation}
    \eta_\mathrm{hom} = \eta_\mathrm{PD} \, \eta_\mathrm{opt} \, \eta_\mathrm{vis} \, \eta_\mathrm{bkg} \, ,
\end{equation}
with $\eta_\mathrm{PD} = \SI{87}{\percent}$ the quantum efficiency of the photodiodes, $\eta_\mathrm{opt} \sim \SI{99}{\percent}$ accounting for optical transmission losses and non-ideal beam splitting, and $\eta_\mathrm{vis} = \mathrm{vis}^2$, where $\mathrm{vis} \geq \SI{94}{\percent}$. 
The factor $\eta_\mathrm{bkg}$ accounts for residual electronic (thermal) noise and laser noise, related to the common-mode rejection ratio (CMRR). 
In our measurements, we obtained $\eta_\mathrm{bkg} \approx \SI{96}{\percent}$, with a CMRR of nearly \SI{54}{\decibel}, achieved through dedicated optimization of the differential amplifier circuit. 
This level of rejection is essential to suppress noise from our commercial laser source.
%
%

\section{Results}\label{sec:results}
\begin{figure}
    \centering
    \includegraphics[width=60mm]{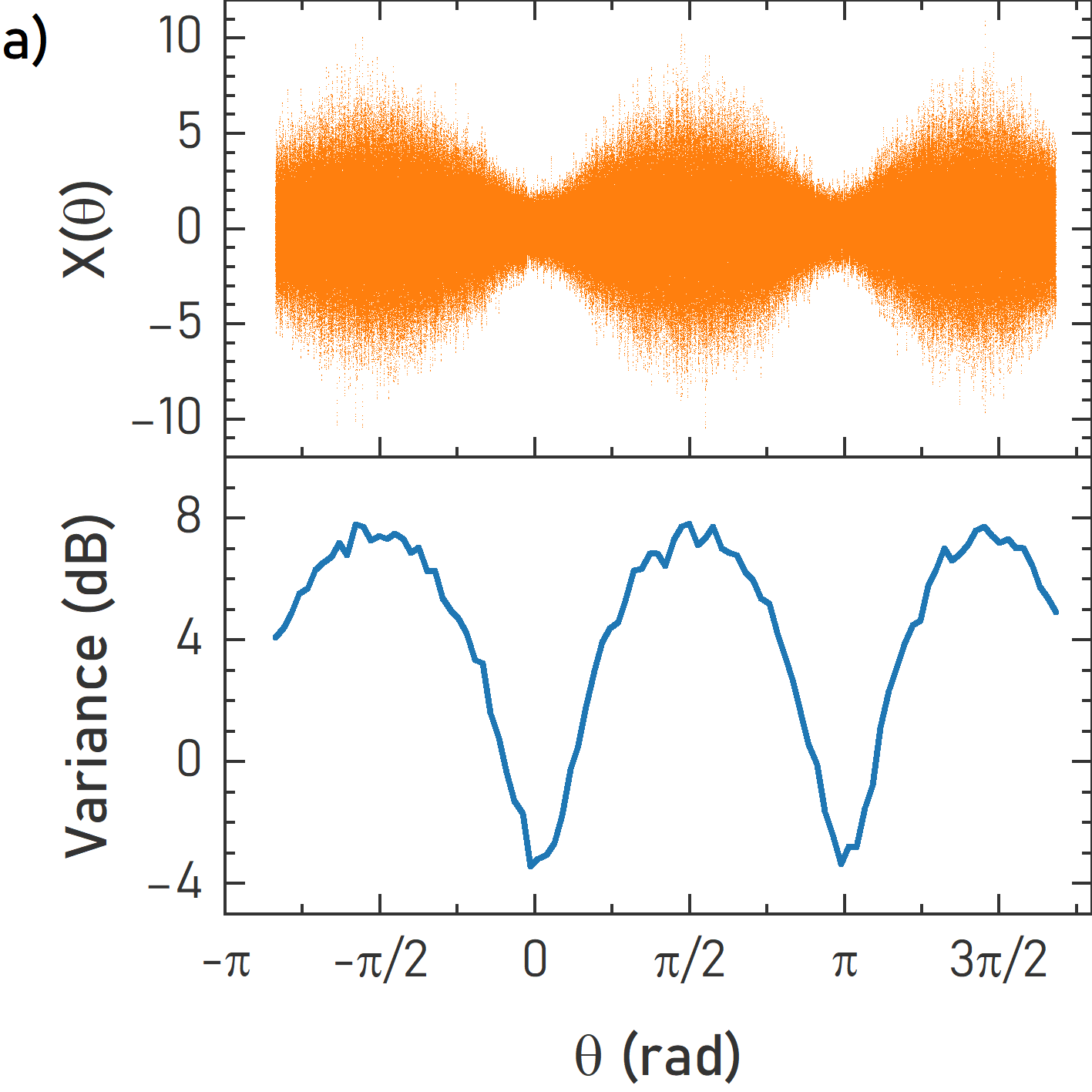}
    \hspace{0.3cm}
    \includegraphics[width=60mm]{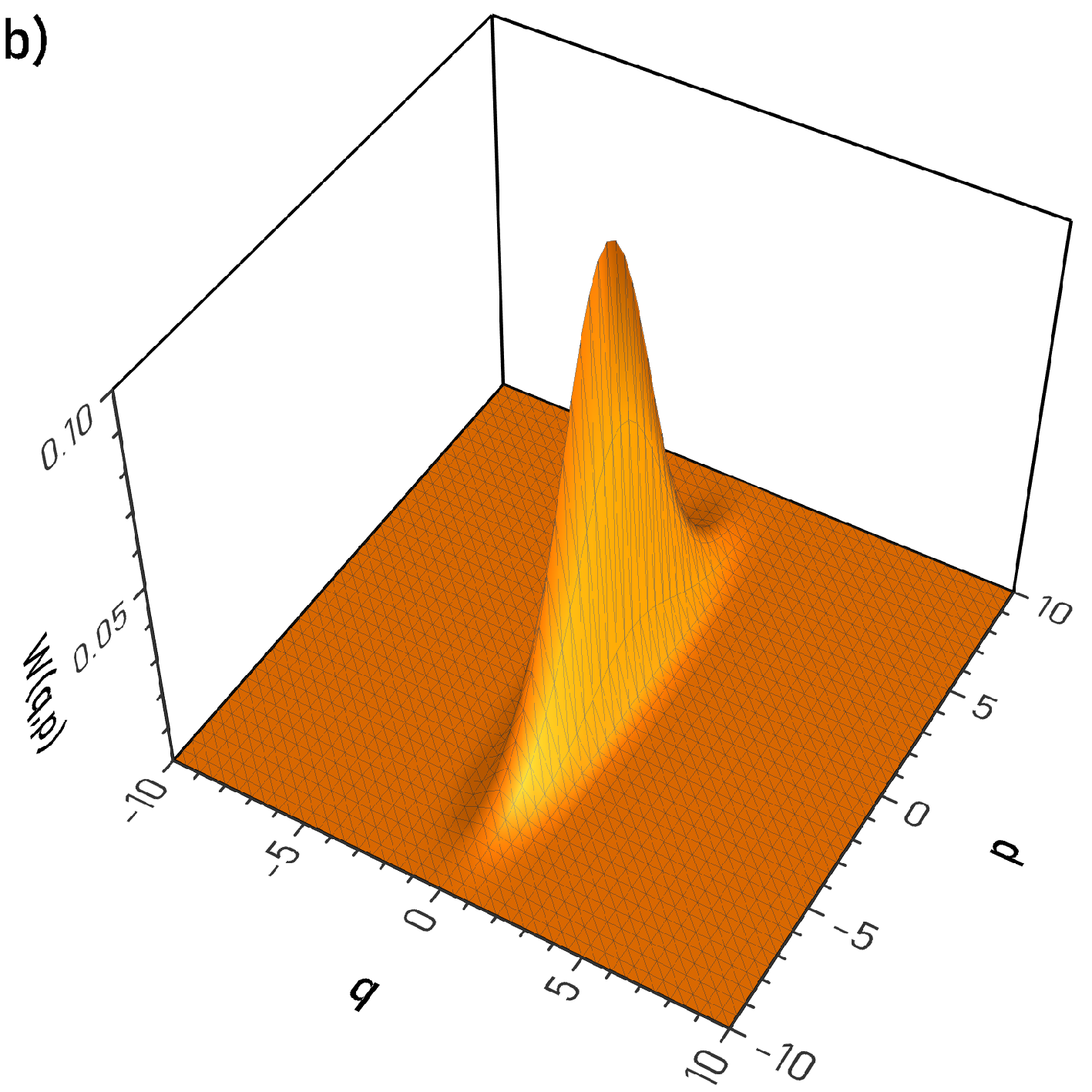}
    \caption{
    a) Measured quadrature noise versus LO phase $\theta$, normalized to the shot-noise level (above), and corresponding squeezing level in \SI{}{\decibel} (below).
    Here, $R_\mathrm{oc}=\SI{81}{\percent}$, and intracavity dispersion is compensated with a chirped mirror of \SI{-900}{\femto\second\squared}.
    b) Wigner function associated with the state outgoing the SPOPO as reconstructed from the experimental homodyne trace in the left panel of the figure and corresponding to a squeezed-thermal state with $\SI{72}{\percent}$ purity and \SI{-5.7}{\decibel} of squeezing (see the text for details).
    }
    \label{fig:tomo}
\end{figure}

As a first step, we validated the operation of the pulsed SPOPO by performing balanced homodyne detection. 
Figure \ref{fig:tomo}a shows an example of the measured quadrature noise as a function of the LO phase, normalized to the shot-noise level, together with the corresponding squeezing level in \SI{}{\decibel}.
Hereafter, we denote by $P_p$ the pump power in \SI{}{\milli\watt}, by $P_\mathrm{th}$ the SPOPO threshold power in \SI{}{\milli\watt}, and by $P = P_p/P_\mathrm{th}$ the pump power normalized to threshold.
This representative measurement was performed with $R_\mathrm{oc} = \SI{81}{\percent}$ and a pump power of $P_p = \SI{40}{\milli\watt}$, corresponding to $P=0.3$, and a parametric gain of $5$.
The visibility of the homodyne detection was \SI{94.7}{\percent}.
For the quadrature noise measurement $X\!\left(\theta\right)$, we acquired \SI{2e6}{} points over a \SI{100}{\milli\second} interval (corresponding to an oscilloscope sampling rate of \SI{20}{\mega Sa \per\second}), covering approximately one full period of the LO phase. 
The variance was computed using a moving average over a window of \SI{20e3}{} points.
The selected acquisition rate and averaging window were verified to operate in a regime where neither the phase-scan speed nor the smoothing affected the observed squeezing level.

Figure \ref{fig:tomo}b shows the Wigner function of the state produced by the SPOPO reconstructed from the experimental data, which, in this example, represents a squeezed-thermal state with quadrature variance (as a function of the quadrature phase) given by \cite{Olivares2012}:
\begin{equation}\label{var:sq}
\sigma_{\rm SPOPO}^2(\theta) = (1+2 n_{\rm th})\left(
e^{-2r} \cos^2\theta +
e^{2r} \sin^2\theta
\right)
\end{equation}
with $n_{\rm th} = 0.20$ and $r=0.83$, corresponding to \SI{-5.74}{\decibel} of squeezing and nonclassical depth ${\cal T} = 0.37$ \cite{Lee1991}. We note that the actual measured variance associated with Equation (\ref{HD:var}) is related to the measured one in Figure \ref{fig:tomo}a by setting $\sigma_{\rm SPOPO}^2(\theta+\alpha \theta^2)$ with $\alpha = 1.25\times 10^{-2}$ in Equation (\ref{var:sq}) to account for the small nonlinearity affecting the experimental phase scan system.

\begin{figure}
    \centering
    \includegraphics[width=60mm]{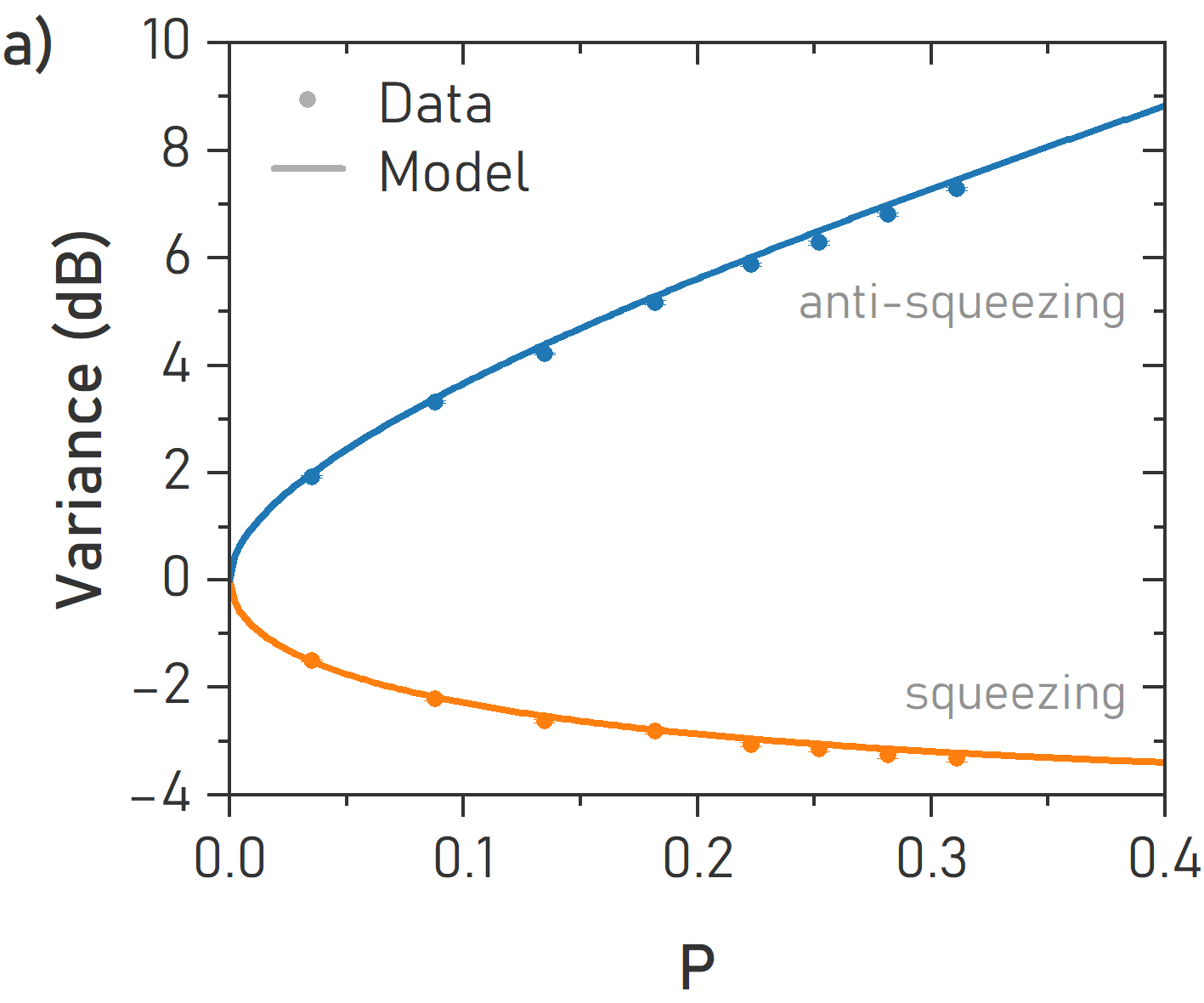}
    \hspace{0.2cm}
    \includegraphics[width=60mm]{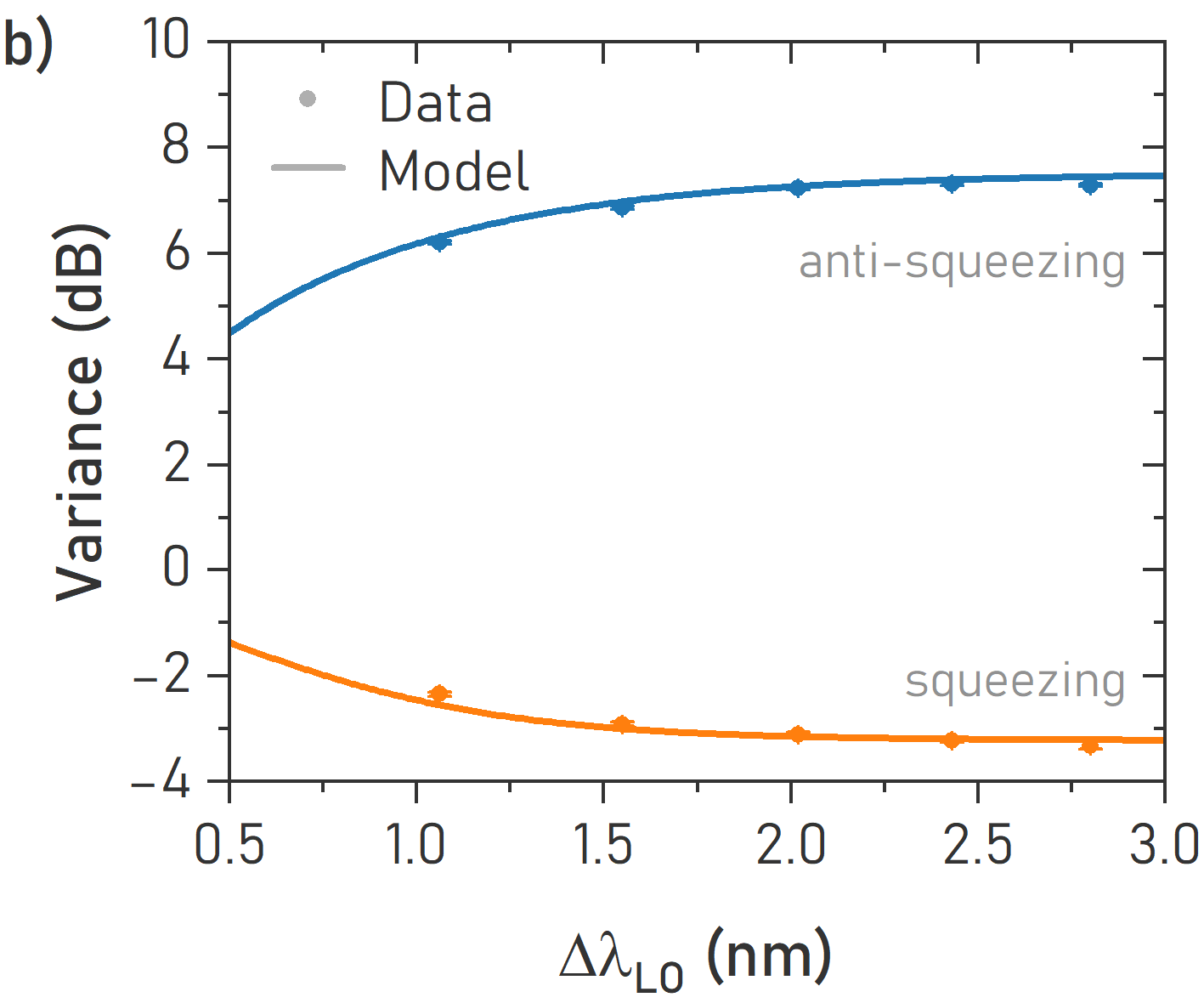}
    \caption{
    a) Squeezing and anti-squeezing levels as a function of pump power normalized to the SPOPO threshold.
    b) Squeezing and anti-squeezing levels as a function of LO spectral width.
    Lines correspond to theoretical predictions, while points are experimental values.
    Error bars ($\sim\pm\SI{0.2}{\decibel}$) are within the marker size.
    }
    \label{fig:squeezing_meas}
\end{figure}

Figure \ref{fig:squeezing_meas}a shows the dependence of the measured squeezing and anti-squeezing levels on the pump power $P$.
As expected, the squeezing increases with pump power, while the anti-squeezing grows more rapidly. 
The solid lines correspond to the predictions of the standard singular-value-decomposition (SVD) model of SPOPOs \cite{Patera2010,MedeirosdeAraujo2014,Roslund2014,Fabre2020}, evaluated with the measured parameters of our setup and corrected for the overall homodyne efficiency $\eta_\mathrm{hom}$. 
In this framework, the variance at the SPOPO output can be expressed as a weighted sum of the contributions of the individual supermodes $k$, according to their overlap with a Gaussian LO: 
\begin{equation}
    \sigma^2_{\mathrm{SPOPO}} = 
    \sum_k \left|M_k\right|^2 \,\sigma^2_{\mathrm{sq},k}
    +
    \left(1 - \sum_k \left|M_k\right|^2\right) ,
\end{equation}
where $\sigma^2_{\mathrm{sq},k}$ denotes the variance of mode $k$ at the SPOPO output (with shot-noise variance set to $1$), and $M_k$ is its overlap with the Gaussian LO. 
The first term accounts for the contribution of the modes projected onto the LO, while the term in parentheses corresponds to the residual unprojected part, i.e., vacuum noise with unit variance. 
In practice, our simulations evaluate the sum up to a finite cutoff $k_0$, beyond which the squeezing level is essentially indistinguishable from vacuum. 
For $k > k_0$, the contribution is therefore taken as vacuum, which prevents numerical artifacts in the simulations. 
The agreement between data and model confirms the validity of this approach.
At $P=0.3$, we measured \SI{-3.3}{\decibel} of detected squeezing, corresponding to \SI{-5.7}{\decibel} at SPOPO output. 
These values are consistent with other similar pulsed multimode SPOPO platforms, such as the $\sim\SI{-6}{\decibel}$ obtained in selected supermodes in Reference \cite{MedeirosdeAraujo2014}.

\begin{figure}
    \centering
    \includegraphics[width=100mm]{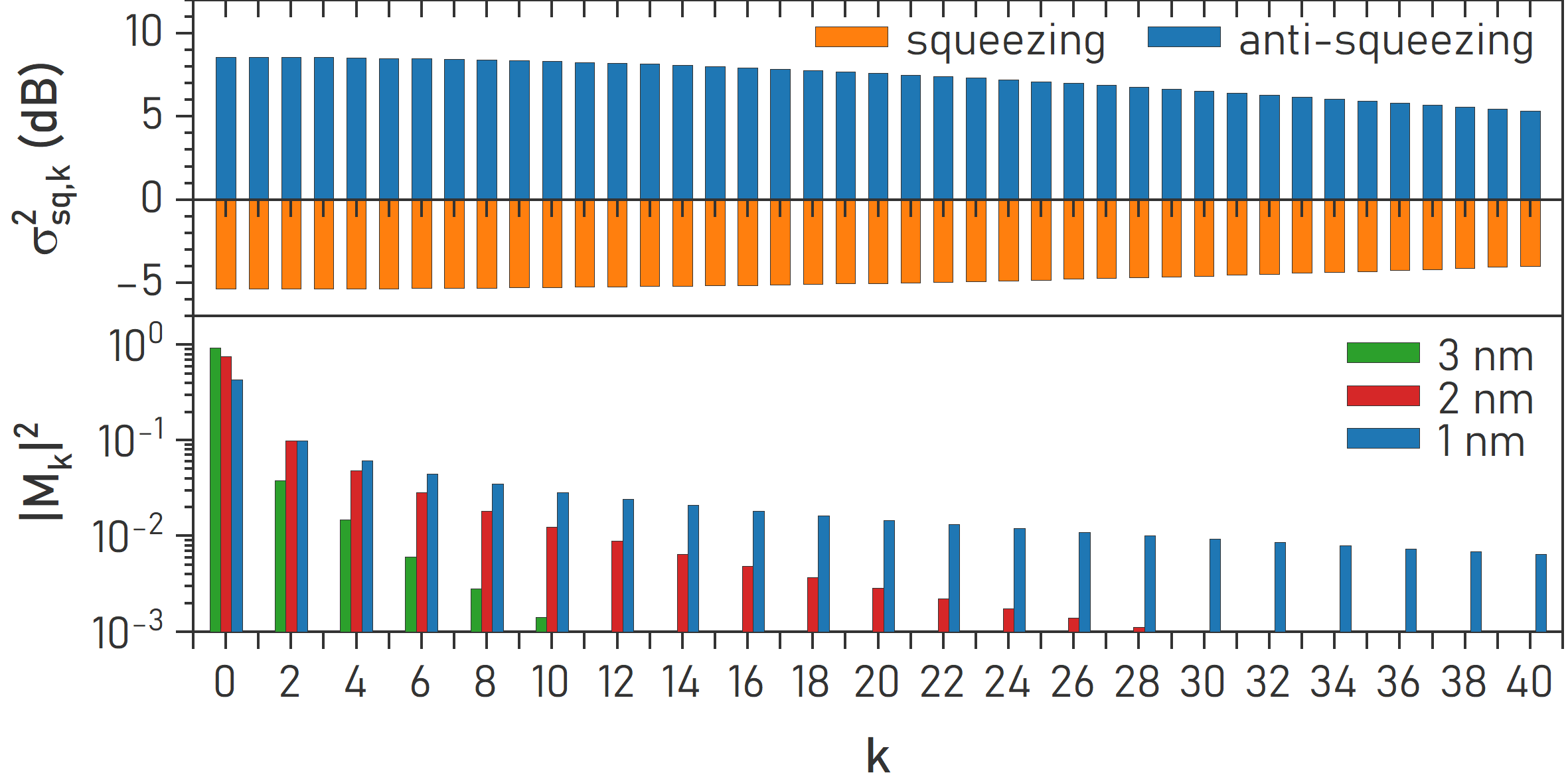}
    \caption{
    Squeezing and anti-squeezing level contribution at the SPOPO output, $\sigma_\mathrm{sq,k}^2$, of the first $40$ supermodes $k$ (top), along with their projection, $\left|M_k\right|^2$, onto LOs with different spectral widths (bottom) of \SI{1.0}{\nano\meter}, \SI{2.0}{\nano\meter}, and \SI{3.0}{\nano\meter} FWHM.
    In this case, the fundamental supermode has a spectral width of \SI{4.4}{\nano\meter}-FWHM.
    }
    \label{fig:mode_projLO}
\end{figure}

Beyond pump power, the same model also predicts how the detected squeezing depends on the LO spectral width. 
We investigated this effect experimentally.
Figure \ref{fig:squeezing_meas}b shows the measured squeezing and anti-squeezing levels as a function of the LO spectral width. 
During these measurements, the homodyne visibility was \SI{94.3}{\percent}.
As the LO bandwidth decreases, both squeezing and anti-squeezing converge toward the vacuum level, in agreement with theoretical predictions.
The explanation of this trend resides in the fact that, as the LO spectrum becomes narrower, its overlap with low-order supermodes decreases. 
Since these modes carry most of the squeezing, a reduction in overlap leads to a decrease of the measurable squeezing, driving the detected noise closer to the vacuum level.
This behavior is consistent with the multimode structure of SPOPO states, where squeezing is distributed over orthogonal supermodes \cite{Patera2010}.
The consistency of this trend is further supported by theoretical simulations shown in Figure \ref{fig:mode_projLO}, which display the contribution to the squeezing level at the SPOPO output, $\sigma_\mathrm{sq,k}^2$, of the first $40$ supermodes $k$ (top), along with their projection, $\left|M_k\right|^2$, onto LOs with different spectral widths (bottom) of \SI{1.0}{\nano\meter}, \SI{2.0}{\nano\meter}, and \SI{3.0}{\nano\meter} FWHM.

Collectively, these results confirm that our setup allows controlled generation and detection of pulsed squeezed states over a range of pump powers and LO spectral widths. 
During the measurements, both the homodyne visibility and the CMRR remained almost constant within experimental uncertainty, confirming the stability of our detection scheme.

\section{Conclusions}\label{sec:conclusions}
We have demonstrated a versatile setup for the generation and detection of pulsed squeezed states using a synchronously pumped optical parametric oscillator. 
Our scheme employs a counter-propagating beam for SPOPO stabilization, which is also used as the local oscillator for homodyne detection, providing a near-perfect mode overlap with the squeezed quantum state. 
The use of spatial light modulators enables independent control of the pump and LO spectra, allowing flexible operation of the system.
Representative squeezing measurements, performed as a function of pump power and LO bandwidth, show good agreement with a multimode SVD-based theoretical model. 
We observed up to \SI{-3.3}{\decibel} of detected squeezing (\SI{-5.7}{\decibel} at SPOPO output), confirming the stability and reliability of the setup.
Such distinctive architecture provides a robust basis for exploring multimode quantum states and implementing advanced protocols in quantum optics and quantum information.

\backmatter


\section*{Declarations}
{\bf Funding} This work has been funded by the Istituto Nazionale di Fisica Nucleare (INFN) within the project T4QC. Sebastiano Corli has been supported by the project QXtreme.\\
{\bf Conflict of interest} The authors declare no competing interests.\\
 {\bf Ethics approval and consent to participate} Not applicable.\\
 {\bf Consent for publication} Not applicable.\\
 {\bf Data availability} The data that support the findings of this study are available from the corresponding author upon reasonable request.\\
 {\bf Materials availability} Not applicable.\\
 {\bf Code availability} Not applicable.\\
 {\bf Author contribution}
Edoardo Suerra: Conceptualization, Data curation, Formal analysis, Investigation, Methodology, Software, Supervision, Validation, Visualization, Writing – original draft, Writing – review \& editing;
Samuele Altilia: Writing – original draft, Writing – review \& editing;
Stefano Olivares: Data curation, Formal analysis, Software, Visualization, Writing – original draft, Writing – review \& editing;
Alessandro Ferraro: Writing – review \& editing;
Francesco Canella: Investigation, Methodology, Validation, Visualization, Writing – original draft, Writing – review \& editing;
Dario Giannotti: Investigation, Methodology, Validation, Visualization, Writing – original draft, Writing – review \& editing;
Gianluca Galzerano: Writing – original draft, Writing – review \& editing;
Sebastiano Corli: Writing – review \& editing;
Enrico Prati: Writing – review \& editing; 
Simone Cialdi: Conceptualization, Data curation, Formal analysis, Funding acquisition, Investigation, Methodology, Project administration, Resources, Software, Supervision, Validation, Visualization, Writing – original draft, Writing – review \& editing.


\bibliography{biblio}

\end{document}